\documentclass[twocolumn,prl,aps,superscriptaddress,showpacs]{revtex4}
\usepackage{graphicx}
\begin{document}

\title{Quaternate  generalization of Pfaffian state at $\nu=5/2$  }

\author{Yue Yu}
\affiliation{Institute of Theoretical Physics, Chinese Academy of
Sciences, P.O. Box 2735, Beijing 100080, China}
\date{\today}
\begin{abstract}
We consider a quaternately generalized Pfaffian
QGPf$(\frac{1}{J(z_i,z_j,z_k,z_l)})[J(z_1,\cdots,z_N)]^2$ in which the
square of Vandermonde determinant, $[J(z_1,\cdots,z_N)]^2$,
implies the upmost Landau level is half filled. This wave function
is the unique highest density zero energy state of a special
 short range interacting Hamiltonian.
One can think this quaternate composite fermion liquid as
a competing ground state of Moore-Read (MR) Pfaffian state at
$\nu=5/2$. The degeneracy of the quasihole excitations above the QGPf is higher than
that of Moore-Read even Read-Rezayi quasiholes. The QGPf is related to a unitary conformal field theory with $Z_2\times Z_2$ parafermions
in coset space $SU(3)_2/U(1)^2$ . Because of the level-rank duality
between $SU(3)_2$ and $SU(2)_3$ in conformal field theory,
these quasiholes above this QGPf state obeying non-abelian
anyonic statistics are expected to support the universal quantum computation at $\nu=5/2$ as Read-Rezayi
quasiholes at $\nu=13/5$. The edge states of QGPf are very different from those of the
Pfaffian's.

\end{abstract}


\maketitle

\noindent {\it Introduction ---} Fractional quantum Hall (FQH) states in
the second Landau level exhibit a very complicated behavior
because of the competition among many nearly degenerate states
\cite{exp1,exp2}. There are two very interesting FQH states with $\nu=\frac{5}2$ and $\frac{12}5$ whose
quasihole excitations are expected to obey the non-abelian
anyonic statistics \cite{mr,rr}, which is thought as the key to open the door to topological
quantum computation \cite{kit,fre}.

The $\nu=\frac{5}2$ FQH state
may possibly be explained by Moore-Read(MR) Pfaffian state \cite{mr}.  It
is two-electron cluster liquid state and can be thought as a
$p_x+ip_y$ weak pairing state of the composite fermions
\cite{gww,gr}. The electron cluster states are not strange in a perpendicular magnetic fields. At higher Landau
levels, say $3<\nu<4$, there are multi-electron bubble crystals
away from the half filling. The electron number within a bubble
may exceed two as the filling factor gradually goes to a half and
a unidirectional charge density wave appears near the half
filling \cite{kfs}. Experimental evidence was already seen
\cite{Lilly}. At $\nu=5/2$, experiments found a transition from
the FQH state to a unidirectional charge
density wave  when an in-plane magnetic field is applied
\cite{inp}.

Although it is widely believed that the FQH
state at $\nu=5/2$ is a $p$-wave weak paired composite fermion
state, none of experimental data confirms it. Moreover, if the
$\nu=13/5$ and 12/5  FQH states are described
by three-electron cluster Read-Rezayi state \cite{rr}, the pairing
state in $\nu=5/2$ seems to be contrary to the appearing order of
number of electrons in a single cluster in higher Landau level in
which the electron number in a single cluster increases as the
filling factor closes to the half filling. The physical origin of
these electron clusters appearing is the same: The lower fully
filled Landau levels' screening supplies an effective attraction
between electrons in the partially filled upmost Landau level.
Therefore, one may raise a question: Is it possible that the
electron number in a cluster exceeds two at $\nu=5/2$?

Numerically, although Morf gave a nice evidence that the MR
Pfaffian state is energetically favored comparing to other
competing states \cite{morf}, only the even total number of electrons
was examined because MR state pairing state is for even number of
electrons. For the odd total electron number, the extra unpairing
electron restricts the boundary condition of the finite system
\cite{gr} and  there is
no numerical study.

It was known that the non-abelian statistics of MR quasiholes is
not sufficiently dense for a universal quantum computer
\cite{fre}. The Read-Rezayi quasiholes are dense to realize the
universal quantum computing but the FQH state
at $\nu=13/5$ (or 12/5) is more delicate. Is it possible that the quasiholes
of the multi-electron clustered state at $\nu=5/2$ provide a base
of the universal quantum computation? This is another motivation
to consider the 4-electron cluster state at $\nu=5/2$.

In this paper, we propose a quaternately generalized Pfaffian (QGPf) state if the
electron number is integer times of 4. The pairing
picture of composite fermions for the Pfaffian is naturally
generalized to quaternate composite fermions. We find that the QGPf
state may be the unique highest density zero energy state of a special Hamiltonian with
short range interaction. The quasihole wave functions
are the higher flux zero energy states of the
special Hamiltonian. The quasihole degeneracy is much higher than that of
MR even Read-Rezayi quasiholes. The QGPf state is related to
a $c=6/5$ unitary conformal field theory(CFT) with $Z_2\times Z_2$ parafermions  in coset space $SU(3)_2/U(1)^2$  \cite{nov}.
Due to the level-rank duality between $SU(3)_2$ and $SU(2)_3$
in CFT, these quasiholes
obeying non-abelian anyonic
statistics are expected to support the universal quantum computation. We also discuss the
edge states of the QGPf and find that they are very different from
those of Pfaffian.

\noindent QGPf {\it Wave Functions ---} We consider two-dimensional
spin polarized electron gas in the second Landau level.
The mixing between Landau levels is neglected and the
second Landau level is treated as the lowest Landau level (LLL)
except the interaction between electrons is renormalized due to
the screening of the electrons in the LLL. We
focus on the half filling, i.e., $\nu=5/2$. For even number of
electrons, we recall the MR Pfaffian state \cite{mr}, e.g., for
8-electrons, which is given by
\begin{eqnarray}
&&{\cal
S}[(z_1-z_3)(z_1-z_4)(z_1-z_5)(z_1-z_6)(z_1-z_7)(z_1-z_8)\nonumber\\
&&(z_2-z_3)(z_2-z_4)(z_2-z_5)(z_2-z_6)(z_2-z_7)(z_2-z_8)\nonumber\\
&&(z_3-z_5)(z_3-z_6)(z_3-z_7)(z_3-z_8)\nonumber\\
&&(z_4-z_5)(z_4-z_6)(z_4-z_7)(z_4-z_8)\nonumber\\
&&(z_5-z_7)(z_5-z_8)(z_6-z_7)(z_6-z_8)]J(z_1,\cdots,z_8)\nonumber\\&&={\rm
Pf}(\frac{1}{z_i-z_j})[J(z_1,\cdots,z_8)]^2
\end{eqnarray}
where ${\cal S}$ denotes the symmetrization of $1,\cdots,8$ and
$J(z_1,\cdots,z_8)=\prod_{i<j\leq 8}(z_i-z_j)$ is the Vandermonde
determinant. Pf$(\frac{1}{z_i-z_j})={\cal
A}(\frac{1}{(z_1-z_2)(z_3-z_4)(z_5-z_6)(z_7-z_8)})$ with ${\cal A}$
denoting the anti-symmetrization. For simplicity, we have omitted
the Guassian factor of the wave function.

The Read-Rezayi state \cite{rr} is a generalization of the MR
Pfaffian. This wave function is for electron number
$3n$ and is a possible competing ground state of spin polarized
electron gas at $\nu=13/5$.
There is another generalization of the MR state to
$N=4n$-electrons ($n>1)$, say $N=8$,
\begin{eqnarray}
&&{\cal S}[(z_1-z_5)(z_1-z_6)(z_1-z_7)(z_1-z_8)
\nonumber\\ &&(z_2-z_5)(z_2-z_6)(z_2-z_7)(z_2-z_8)\label{6wf}
\\ &&(z_3-z_5)(z_3-z_6)(z_3-z_7)(z_3-z_8)
\nonumber\\ &&(z_4-z_5)(z_4-z_6)(z_4-z_7)(z_4-z_8)]
J(z_1,\cdots,z_8).\nonumber
\end{eqnarray}
The symmetrizing part of this wave
function allows 4-particles occurring the same position but not
5-particles. The Vandermonde determinant $J(z_1,\cdots,z_8)$ leads
to the total wave function is anti-symmetric and electrons obey
Pauli principle. This state may be rewritten as
$
{\rm QGPf}(\frac{1}{
J(z_i,z_j,z_k,z_l)})[J(z_1,\cdots,z_8)]^2
$
where the generalized Pfaffian (QGPf) is defined by ${\cal
A}(\frac{1}{ J(z_1,z_2,z_3,z_4)J(z_5,z_6,z_7,z_8)})$ and
$J(z_i,z_j,z_k,z_l)$ is the Vandermonde determinant for $z_i,z_j,
z_k$ and $z_l$. Generalizating to the system with $N=4n$ ($n>1$) electrons, we
have
\begin{eqnarray}
{\rm
QGPf}\biggl(\frac{1}{J(z_i,z_j,z_k,z_l)}\biggr)[J(z_1,\cdots,z_N)]^2,
\label{N}
\end{eqnarray}
where the QGPf is defined by ${\cal
A}(\frac{1}{J(z_1,z_2,z_3,z_4)}
\cdots\frac{1}{J(z_{N-3},z_{N-2},z_{N-1},z_{N})}).$
The filling
factor of this state is, $\nu=\frac{N}{N_\phi}$ for
$N_\phi=2(N-1)-3$, which tends to 1/2 as $N\to\infty$ and
coincides with $\nu=2+1/2$ in the second Landau level.

\noindent{\it Special Hamiltonian ---}  It was known that the MR
Pfaffian state is the highest density zero energy state of the
Hamiltonian $H_{MR}=V\sum_{i<j<k}\delta'(z_i-z_j)\delta'(z_j-z_k)$
\cite{gww}. The Hamiltonian, whose highest density zero energy state
is the Read-Rezayi state, is given by $
H_{RR}=V\sum_{i<j<k<l}\delta'(z_i-z_j)\delta'(z_j-z_k)\delta'(z_k-z_l).$
 Therefore, the MR Pfaffian state and Read-Rezayi state are
the corresponding ground states of these special Hamiltonians \cite{rr}, according to Haldane's highest density criteria
\cite{haldane}.

The QGPf is zero energy state of $H_{RR}$ but not
the highest density one. It is not zero energy state of $H_{MR}$.
Can the QGPf state be a ground state of a special short range
interacting Hamiltonian?  For $N=4n$ electrons, we consider the
following Hamiltonian
\begin{eqnarray}
H&=&V~\sum_{P_{4n}}[\delta'(z_{i_1}-z_{i_2})\delta'(z_{i_2}-z_{i_3})]
\cdots\nonumber\\&&
[\delta'(z_{i_{4n-3}}-z_{i_{4n-2}})\delta'(z_{i_{4n-2}}-z_{i_{4n-1}})]
\nonumber\\&&\delta'(z_{i_{4a}}-z_{i_{4b}}),\label{gpfh}
\end{eqnarray}
where $P_{4n}$ is a permutation of $1,\cdots,4n$ with $i_1<i_2<i_3;\cdots;i_{4n-3}<i_{4n-2}
<i_{4n-1}$, and $i_{4a}< i_{4b}$
with $a\ne b\leq n$.
Here, we divide electrons into $n$-groups with 4 electrons in
each group. Take three in each group and let them interacting with a three-body
short range potential. Left electrons belong to the
distinct groups and the last pair in the Hamiltonian comes from
them. Then make all these electrons interacting simultaneously.
This $H$ outwardly is a $3n+2$-electron interaction and in principle can be
treated by means of the method developed in a recent work by Simon
et al \cite{simon} but it is hard to handle when electron number
becomes large. However, since it has been fractionized to independent $n$-three-body
interaction and a two-body interaction, this special form of the
interaction here in fact reflects the three-body interaction physics
and it helps us to attract the lowest flux (i.e., the
highest density) zero energy state. Taking $N=8$ as an example,
the Hamiltonian is given by
\begin{eqnarray}
&&H_8=V~[\delta'(z_1-z_2)\delta'(z_2-z_3)\delta'(z_5-z_6)\delta'(z_6-z_7)\nonumber\\&&\delta'(z_4-z_8)
+{\rm other~terms~by~cycling~(1\cdots8)]},\label{6h}
\end{eqnarray}
The zero energy wave function is written as $
\Psi_{electron}(z_1,\cdots,z_8)=\Psi_{symm}(z_1,\cdots,z_8)J(z_1,\cdots,z_8).
$ If we consider only the symmetric part $\Psi_{symm}$, the
$\delta'$-function should be replaced by the $\delta$-function. In
order to find the lowest flux zero energy state, we divide eight
electrons into two groups, say, (1234) and (5678). The relevant
terms in the Hamiltonian are the terms including an inter group
pair, say the pair $z_4-z_8$ in the first term of $H_8$.
Thus, the most economic way
to get the lowest flux
is to include only this pair in $\Psi_{symm}$ which then includes a term
$\prod_{i,j=1}^4 (z_i-z_{4+j})$. One can check that all
other terms in (\ref{6h}) (for $\delta$-function)
act on it vanishing and if taking away
any factor from it, one can always have a non-zero acting.
Therefore, this is a lowest
flux zero energy state. When regrouping the electrons,
the lowest flux state also changes correspondingly. Due to the
total symmetry of $\Psi_{symm}$, regrouping leads to a unique lowest flux state,
 i.e.,
$
\Psi_{symm}(z_1,\cdots,z_8)={\cal S}[\prod_{i,j=1}^4
(z_i-z_{4+j})],
$
which is exactly the symmetric factor in (\ref{6wf}).
This is the unique lowest flux zero energy state of $H_8$.
This argument for eight electrons is also true for arbitrary $4n$
electrons because one can always think each term in the
Hamiltonian (\ref{gpfh}) contains only one inter group electron
pair. Therefore, the QGPf wave function is the ground state of the
special Hamiltonian (\ref{gpfh}).
 The MR Pfaffian is also the zero energy state but
has a higher flux.

\noindent{\it Quasiholes ---} Since the wave function must be totally
antisymmetric, similar to MR quasiholes in pairs\cite{mr,nw}, the quasiholes
create in quaternions, e.g., the
$4$-quasihole wave function is given by
\begin{eqnarray}
{\rm
QGPf}\biggl(\frac{f(z_i,z_j,z_k,z_l;w_1,w_2,w_3,w_4)}{J(z_i,z_j,z_k,z_l)}\biggr)
,\label{quasi}
\end{eqnarray}
where
$
f(z_i,z_j,z_k,k_l;w_1,w_2,w_3,w_4)
=(z_i-w_1)(z_j-w_2)(z_k-w_3)(z_l-w_4)+(ijkl)~{\rm cycle}$.
This is a zero energy state of the Hamiltonian (\ref{gpfh}) with the flux increasing to
$N_\phi=2N-2$ \cite{nick}. Note that if $w_1=w_2=w_3=w_4$, it gives a Laughlin
quasihole with charge $1/2$. Therefore, the quasihole charge is
$1/8$. The $4m$-quasihole wave function can be defined in a similar way with
\begin{eqnarray}
&&f(z_i,z_j,z_k,z_l;w_1,\cdots,w_{4m})
\nonumber\\
&&=(z_i-w_1)\cdots(z_i-w_m)(z_j-w_{m+1})\cdots(z_j-w_{2m})\nonumber\\
&&\times(z_k-w_{2m+1})\cdots (z_k-w_{3m})\nonumber\\
&&\times(z_l-w_{3m+1})\cdots (z_l-w_{4m})+(ijkl)~{\rm cycle}. \label{quasig}
\end{eqnarray}
By exchanging the coordinates of quasiholes among four different
sets $(w_{am+1},\cdots,w_{(a+1)m})$ ($a=0,1,2,3$), we can have
$C^{4m-1}_{m-1}C^{3m-1}_{m-1}C^{2m-1}_{m-1}$ states. However, these quasihole
states are not all independent. For example, for $m=2$, we have eight quasiholes and
35 different quasihole wave functions. A key relation to pick out the independent
states reads \cite{nw}
\begin{eqnarray}
[12]_i[34]_j-[14]_i[23]_j=x([12]_i[34]_j-[13]_i[24]_j),
\end{eqnarray}
where  $[12]_i=(z_i-w_1)(z_i-w_2)$,
$x=\frac{w_{13}w_{24}}{w_{14}w_{23}}$ and $w_{12}=w_1-w_2$, etc.
If we fix 4 in 8 $w_i$,
one can follow Ref. \cite{nw} step by step
to check this relation is also correct
when it is put into the QGPf. This means we have 3
different wave functions and only 2 of them
are independent. If we fix two $w_i$, there are 15 different wave functions
and 5-indenpendent ones. For 35 different
wave functions of total 8 quasiholes, there are 13 linearly independent. In general, according to
exclusion statistics point of view \cite{shouten},
this degeneracy is equal to a generalized Finobacci number $F^g_{4m-3}$ which is
defined by $F^g_n=F^g_{n-1}+F^g_{n-2}+F^g_{n-3}$ with $F^g_0=F^g_1=1$ and$F^g_2=F^g_1+F^g_0$.
$F^g_1=1$ and $F^g_5=13$ is consistent with $m=1$ and $m=2$ calculations.

There are three kinds of twisted states: (1) Take $w_1=0$ and
$w_2=w_3=w_4=\infty$ in the three quasihole wave function;
(2) Take $w_1=w_2=0$ and $w_3=w_4=\infty$; (3) Take $w_1=w_2=w_3=0$
and $w_4=\infty$.

\noindent{\it Conformal Field Theory ---} A unitary CFT
may related to the QGPf state is a $Z_2\times Z_2$ parafermion theory
with coset space $SU(3)_2/U(1)^2$ and $c=\frac{kD}{k+g}-2=6/5$ ($D=8,k=2,g=3$) \cite{nov}.
The $Z_3$ parafermion theory with coset space $SU(2)_3/U(1)$ supports
the universal quantum computation \cite{fre}. Because of the level-rank duality between $SU(3)_2$ and
$SU(2)_3$, we expect the parafermion theory in coset space $SU(3)_2/U(1)^2$ also supports
the universal quantum computation.

There are three Majorana fermions $\psi^\alpha$, $\alpha=1,2,3$, which are parafermions graded by
$Z_2\times Z_2$ (with the identity). The OPEs are given by
\begin{eqnarray}
&&\psi^\alpha(z)\psi^\alpha(w)=\frac{1}{z-w}+O((z-w)^0) \\
&&\psi^\alpha(z)\psi^\beta(w)
=\frac{c_{\alpha\beta}\psi^\gamma(w)}{(z-w)^{1/2}}
+O((z-w)^{1/2}),\nonumber
\end{eqnarray}
where $\alpha\ne\beta\ne\gamma$ in the second equation and
$c_{12}=c_{23}=c_{31}=e^{-i\pi/4}/\sqrt{2}$ and $c_{\beta\alpha}=c_{\alpha\beta}^*$.
Three point parafermion correlation function for $\alpha\ne \beta\ne \gamma$ is give by \cite{nov}
\begin{eqnarray}
\langle\psi^\alpha(z_1)\psi^\beta(z_2)\psi^\gamma(z_3)\rangle
=\frac{e^{-i\epsilon^{\alpha\beta\gamma}\pi/4}}
{z_{12}^{1/2}z_{13}^{1/2}z^{1/2}_{23}} \label{paraexp}
\end{eqnarray}
Three twisted primary  fields $\sigma^{12},\sigma^{23},\sigma^{13}$ have conformal
dimension 1/10.
When acting $\psi^\alpha $ to
$\sigma^{\alpha\beta}$, it behaves like
the Ising spin field $\sigma^\alpha$.  Using the OPEs, one has
\begin{eqnarray}
{\cal A}\{\langle{\cal N}[\prod_{\alpha=1}^3(:\prod_{i=1}^4\psi^\alpha(z_i):)]\rangle\}
=\frac{1}{J(z_1,\cdots,z_4)},
\end{eqnarray}
where ${\cal N}$ is defined by subtracting the singularity from, e.g.,
$\psi^1(z_1)\psi^2(z_1)$, etc. That is, let all $\psi^2=\psi^2(z+\epsilon)$ and
$\psi^3=\psi^3(z+2\epsilon)$. Then subtract the divergence with a term $O(1/\epsilon^{1/2})$ and
take $\epsilon\to 0$ at the end of calculations.  The normal ordering is $:\psi^\alpha(z)\psi^\alpha(w):
=\psi^\alpha(z)\psi^\alpha(w)-\frac{1}{z-w}$, and so on. That is, we forbid the direct contraction between
the same type Majorana fermions. Notice that this result is a branch cut-free generalization to (\ref{paraexp}).
To get the QGPf, we calculate
\begin{eqnarray}
&&{\cal A}
\{\langle{\cal N}[ \prod_{\alpha=1}^3(:\prod_{i=1}^N\psi^\alpha(z_i):)]\rangle\}
={\rm QGPf}(G_4)\\&&
+{\cal A}[G_4^{N-8}G_3G_5+G_4^{N-12}(G_5G_7+G_3G_9)+\cdots],\nonumber
\end{eqnarray}
where $G_3=1/J(z_i,z_j,z_k);
G^s_4=1/[J(z_{i_1},z_{i_2},z_{i_3},z_{i_4})\cdots
J(z_{i_{s-3}},z_{i_{s-2}},z_{i_{s-1}},z_{i_s})];$ $ G_{2a+1}(z_1,\cdots,z_{2a+1})
=1/[z_{12}z_{23}\cdots z_{2a+1,1}z^{1/2}_{13}
\cdots z^{1/2}_{2a,1}z^{1/2}_{2a+1,2}]$ for $2a+1\geq 5$
. All terms in square brackets on the right side include
branch cut factors which can not be cancelled by multiplying a Jastraw factor $[J(z_1,\cdots,
z_N)]^{p/q}$.
Projecting them away implies projecting to the LLL. Therefore, the
lowest Landau level projection leaves  the QGPf only.

We notice that not all quasihole wave functions (\ref{quasig}) can be fallen
under a correlations function
of this CFT. Using the twisted primary field
what we can get is the following correlation function:
\begin{eqnarray}
{\cal A}
\{\langle\sigma^{12}(w_1)\sigma^{12}(w_2)\sigma^{12}(w_3)
\sigma^{12}(w_4){\cal N}[\prod_{\alpha=1}^3:\prod_{i=1}^N\psi^\alpha(z_i):]
\rangle\}_{LLL}
\nonumber\\\sim {\rm QGPf}(\frac{f_4(z_i,z_j,z_k,z_l;w_1,w_2,w_3,w_4)}
{J(z_i,z_j,z_k,z_l)}),~~~~~~~~~~~\nonumber
\end{eqnarray}
where $f_4=(z_i-w_1)^4(z_j-w_2)^4(z_k-w_3)^4(z_l-w_4)^4+(ijkl)~{\rm cycle}$.
This is a 16 quasihole wave function with four at the same position.

\noindent{\it Edge Excitations --} The edge excitations can also
be discussed by a parallel way to those in the Pfaffian state
\cite{edge}. The Laughlin-type charge edge excitations are exactly
the same as those in the Pfaffian state, which can be obtained by
timing symmetric polynomials to the QGPf state. However, the
neutral edge excitations are very different from those of the
Pfaffian state, which are given by replacing the QGPf state by
\begin{eqnarray}
{\cal A}(z_1^{p_1}\cdots
z_{4m'}^{p_{4m'}}\frac{1}{J(z_{4m'+1},z_{4m'+2},z_{4m'+3},z_{4m'+4})}\cdots)
\end{eqnarray}
These edge states gain the momentum $\Delta
M=\sum_{i=1}^{4m'}(p_i+3/2)$, instead of $\sum_i (p_i+1/2)$ for Majorana fermions.

For twisted states, the edge excitations are given by
$
{\cal A}(z_1^{p_1}\cdots z_{4m'}^{p_{4m'}}
\frac{z_{4m'+1}+z_{4m'+2}+z_{4m'+3}+z_{4m'+4}}{J(z_{4m'+1},z_{4m'+2},z_{4m'+3},z_{4m'+4})}\cdots)
$
with $\Delta M$ $=\sum_{i=1}^{4m'}(p_i+13/8)$  and
other two raise momentum $\Delta M=\sum_{i=1}^{4m'}(p_i+7/4)$ and
 $\Delta M=\sum_{i=1}^{4m'}(p_i+15/8)$ .

\noindent{\it Experimental Implication --} Experimentally, the
charge of the quasiparticle may be measured by the shot noise in a
point contact tunnelling experiment, as measuring the fractional
charge of the Laughlin quasiparticle \cite{charge}. For the MR
Pfaffian state, the quasihole charge is $\frac{e}4$ while it is
$\frac{e}8$ for the QGPf state. Recent proposed
quasiparticle interferometry may measure the non-abelian
statistics of the quasiparticles \cite{stat}. The different
non-abelian statistical property will be reflected in this kind of
experiments.

\noindent{\it Conclusions ---} We have constructed
a competing wave function of four-electron cluster in $\nu=5/2$ ,
the quaternate generalization of the pairing of composite fermions.
This incompressible liquid state may challenge the MR Pfaffian
state. The corresponding special Hamiltonian and the CFT
were studied.
The conformal field related to this QGPf state is dual to that of the
Read-Rezayi quasiholes at $\nu=13/5$. Therefore, we expect a universal quantum computation
in $\nu=5/2$. The finite electron calculations
with powerful computational methods are definitely
required to compare with the MR Pfaffian. Because the system is
particle-hole symmetric for the Landau level mixing is
neglected, an anti-QGPf state is expected like the anti-Pfaffian state
competing to the Pfaffian state \cite{ap}.

The author thanks Boris Noyvert, Zhenghan Wang, Xiao-Gang Wen and Zhongyuan Zhu
for useful discussions. This work was supported in part by Chinese NNSF, the national program for basic research of
MOST of China and a fund from CAS.

\noindent{\it Note added} The wave function (\ref{N}) may be a member
of a class of possible FQH wave functions recently proposed by Wen and Wang \cite{wenwang}.

\end{document}